\documentclass[a4paper,fleqn,usenatbib]{mnras}
\usepackage[T1]{fontenc}
\usepackage{ae,aecompl}
\usepackage{graphicx}
\usepackage{amssymb}
\usepackage{amsmath}
\usepackage{mathptmx}
\usepackage{epstopdf}
\usepackage{booktabs}
\usepackage{float}
\usepackage{subfig}
\usepackage{epsfig,color,ulem}
\usepackage{tabularx}

\definecolor{green}{rgb}{0.0, 0.5, 0.0}

\newcommand{\msun}{\,{\rm M_{\odot}}}
\newcommand{\cm}{\,{\rm cm}}
\newcommand{\rad}{\,{\rm rad}}
\newcommand{\s}{\,{\rm s}}
\def\gsim{ \lower .75ex \hbox{$\sim$} \llap{\raise .27ex \hbox{$>$}} }
\def\lsim{ \lower .75ex\hbox{$\sim$} \llap{\raise .27ex \hbox{$<$}} }

\title[]{
From $\gamma$ to Radio  - The Electromagnetic Counterpart of GW 170817}

\author[ Nakar et al., ]{
	Ehud Nakar$^{1}$\thanks{udini@wise.tau.ac.il},
	Ore Gottlieb$^{1}$
	Tsvi Piran$^{2}$
	Mansi M. Kasliwal$^{3}$
	Greg Hallinan$^{3}$
	\\
	$^{1}${The Raymond and Beverly Sackler School of Physics and
		Astronomy, Tel Aviv University, Tel Aviv 69978, Israel}\\
	$^{2}${Racah Institute of Physics, The Hebrew University of
		Jerusalem, Jerusalem 91904, Israel}\\
		$^{3}${Division of Physics, Math and Astronomy, California Institute of Technology,
1200 East California Boulevard, Pasadena, CA 91125, USA}\\
}
\pubyear{2018}
\begin{document}
	\label{firstpage}
	\pagerange{\pageref{firstpage}--\pageref{lastpage}}
	\maketitle	
\begin{abstract}
The gravitational waves from the first binary neutron star merger, GW170817,  
were accompanied by a multi-wavelength electromagnetic counterpart, from $\gamma$-rays to radio. The accompanying gamma-rays, seems at first to confirm the association of mergers with short gamma-ray bursts (sGRBs). The common interpretation was that we see an emission from an sGRB jet seen off-axis. However, a  closer  examination of  the sub-luminous $\gamma$-rays and the  peculiar radio afterglow were inconsistent with this simple interpretation.
Here we present results of 3D and 2D numerical simulations that follow the hydrodynamics and emission of the outflow from a neutron star merger form its ejection and up to its deceleration by the circum-merger medium. Our results show that the entire set of $\gamma$-rays, X-rays and radio observations can be explained by the emission from a mildly relativistic cocoon material (Lorentz factor $\sim$2-5) that was formed while a jet propagated through the  material ejected during the merger. The $\gamma$-rays are generated when the cocoon breaks out from the engulfing ejecta  
while the afterglow is produced by interaction of the cocoon matter with the interstellar medium. The strong early uv/optical signal may be a Lorentz boosted macronova/kilonova.
The fate of the jet itself is currently unknown, but our full-EM models define a path to resolving between successful and choked jet scenarios, outputting coupled predictions for the image size, morphology, observed time-dependent polarization and light curve behavior from radio to X-ray. The predictive power of these models will prove key in interpreting the on-going multi-faceted observations of this unprecedented event. 
\end{abstract}

\section{Introduction} 
At first glance GRB 170817A looks like a regular sGRB  \citep{Goldstein17,savchenko2017}, thereby confirming the association of mergers with short gamma-ray bursts (sGRBs) \citep[e.g.,][]{Eichler1989,nakar2007}. However, a closer examination reveals that it does not resemble any other burst seen before. Its total isotropic equivalent luminosity is smaller by several orders of magnitude than the weakest short burst with an identified redshift. It is softer than typical sGRBs and it has a unique spectral evolution: a  harder ($E_p \sim 185 \pm 62$ keV) $0.6$ sec pulse, followed by a thermal tail ($kT \sim$ 10 keV) lasting about one sec.  
The observed afterglow was not less puzzling.  A bright declining X-ray signal is observed in regular sGRBs from day one. Here, only upper limits were obtained during the first week.  The first X-ray detection \citep{Troja17,Margutti17} was only at day nine. The first radio detection  \citep{Hallinan17} was only at day sixteen.   
 
The event was also accompanied by a uv/optical/IR signal detected 11hr after the merger \citep[e.g.,][]{coulter2017,arcavi2017,soares-santos2017,diaz2017,mccully2017,buckley2018,utsumi2017,covino2017,shappee2017}.  The common  macronova (a.k.a. kilonova) interpretation   is that it was powered by radioactive decay within the dynamical ejecta and winds ejected during the merger  \citep[e.g.,][]{Evans17,kasen2017,cowperthwaite2017,smartt2017,Kasliwal17,Pian17,tanvir2017,tanaka2017,drout2017,kilpatrick2017,nicholl2017,waxman2017,shappee2017}. 
However, even this signal deviated somewhat from the expectations. It was too bright and  too blue  half a day after the merger. While clearly inconsistent with lanthanides high opacity ($\kappa \approx 10 {\rm~cm^2/g}$), even a much lower opacity  ($\kappa \approx 1 {\rm~cm^2/g}$) is hardly consistent with such a bright early signal \citep{Kasliwal17,arcavi2018}. 

The common explanation to the peculiar observations was  that we have observed a regular sGRB jet off-axis    \citep{Troja17,Margutti17,Haggard2017,Goldstein17}. However,  if that was the case, an observer that looked at the jet on-axis would have seen a very a-typical sGRB \citep{Kasliwal17,Granot17}\footnote{Simple Doppler arguments suggest that the peak photon energy of that burst would have been at least $\approx 5$ MeV. Such peak energy is  much higher than the typically observed SGRB  peak energy, and   is seen  only rarely.}.  
{More importantly, the very strong dependence of the observed off-axis luminosity on the observing angle implies that in this interpretation the angle to the gamma-ray emitting region must be small ($\lesssim 0.1$ rad; \citealt{Kasliwal17}). Therefore a clear prediction of this model is a radio and X-ray afterglow that rises to a peak soon after the merger, which is inconsistent with the continuous rise seen during the first $\sim 100$ day \citep{mooley2018,Ruan2018,pooley2017,margutti2018,lyman2018}.}. 
Alternatively, the unique properties of the $\gamma$-rays suggest that we 
observed a mildly relativistic shock breakout. Compactness arguments show that the radiating material must have at least   $\Gamma \approx 2-3$ \citep{Gottlieb17b}, while the uv/optical/IR indicate that $\sim 0.05 {\rm~M_\odot}$  were launched at velocities of $0.1-0.3$c during the merger.
This suggests that a  jet with  $\Gamma \gtrsim 3$ was launched following the merger into the expanding ejecta. {When the jet propagates through the sub-relativistic ejecta it inflates a cocoon \citep{ramirez-ruiz2002}. \cite{Nakar2017}, \cite{lazzati2017a} and \cite{Gottlieb17a} have  shown before GW170817 that in the conditions expected following a neutron star merger the jet propagation  gives rise to a wide-angle mildly relativistic cocoon . Moreover, the cocoon is expected not only in case that the jet successfully breaks out of the sub-relativistic ejecta, presumably producing a regular sGRB for an on-axis observer, but also in cases when the jet is choked within the ejecta and no sGRB is generated.}

Soon after the detection of GW170817 we and others suggested that the $\gamma$-rays were generated by the breakout of the shock driven by the cocoon out of the sub-relativistic ejecta \citep{Kasliwal17,Gottlieb17b,Bromberg2018,Pozanenko2018}. 
The mildly relativistic shock breakout model is appealing since it explains all the properties of the signal: its low luminosity compared to the total explosion energy,  its spectrum, duration and the delay compared to the GW signal. Moreover, the quasi-thermal spectrum and its hard to soft evolution is a clear prediction of shock breakout emission  \citep{nakar2012,Gottlieb17b}. In addition, all the unique properties of GRB 170817A are seen also in low-luminosity GRBs (a special family of long GRBs), which are believed to arise from a shock breakout from the stellar surface  
\citep{kulkarni1998,campana2006,nakar2012,nakar2015}.
For a shock breakout to explain the observed $\gamma$-rays in GRB 170817A it should take place at a radius of $10^{11}-10^{12}$ cm  with $\Gamma \sim 3$.

While the $\gamma$-rays gave the first hint for a significant mildly relativistic outflow, the radio and X-ray afterglow provided a very strong support for this picture. 
The radio signal, which seems as the mildly relativistic part of the radio flare predicted by \cite{nakar2011}, rises continuously, roughly as $t^{0.8}$, since the first detection on day sixteen  \citep{Hallinan17,mooley2018} and so does the X-ray signal \citep{Ruan2018,pooley2017,margutti2018}. This moderate continuous increase is inconsistent with any type of  off-axis emission (i.e., emission from material that moves with a Lorentz factor $\Gamma$ at an angle larger than $1/\Gamma$ with respect to observer). Instead it requires a mildly relativistic blast wave with $\Gamma \sim 2-5$ that radiates on-axis \citep{nakar2018a}. Namely, the blast wave propagates  into the ISM at a direction  within an angle of $1/\Gamma$ with respect to the line of sight. 
The continuous rise in the radio implies that between day 10 and 100  the energy in the region that radiates on-axis increased by about a factor of 10. The additional energy can be produced by a slower inner matter which is moving behind the blast wave and is catching up as it slows down or from matter moving on a slightly larger viewing angles that decelerates and comes into the line of sight (i.e., an angle $<1/\Gamma$)  \citep{nakar2018a}. All these properties of the emitting blast wave fit very well the outcome of the interaction of the mildly relativistic cocoon (with either a choked or a successful jet) with the circum-merger inter-stellar medium  \citep{Gottlieb17b,lazzati2017c,margutti2018}. 

The first goal of this paper is to find if a single model can explain the entire electromagnetic observations, starting with the prompt gamma-rays, through the uv/optical/IR macronova and up to the radio and X-ray afterglow. The second goal is to find out what future observation are needed to distinguish between different models that currently fit the data, and specifically whether there is a successful jet in GW170817 or not. For that we carry out a series of numerical simulations which are the first to encompass the entire  evolution and emission, from the prompt $\gamma$-rays to the afterglow, in a single simulation. A general description of the simulations is given in \S  \ref{sec:methods}. In order to streamline the structure of the paper the exact details of the simulations are discussed in an Appendix. We discuss the resulting light curves of the $\gamma$-rays, the uv/optical/IR macronova and the radio afterglow in \S  \ref{sec:results}. In  \S \ref{sec:jet} we explore the differences in the predicted radio light curve, image and polarization between a choked and a successful jet and discuss how future observations may distinguish between these two scenarios.
 We summarize our results and conclude in \S \ref{sec:summary}.

\section{Methods} 
\label{sec:methods} 
We have carried out a set of relativistic hydrodynamic simulations (see Appendix for details) of the post-merger outflow starting at the jet launching, following its propagation through the sub-relativistic ejecta and the cocoon shock breakout, where the $\gamma$-rays are produced, continuing to the homologous phase during which the uv/optical/IR macronova is emitted and ending in the interaction of the outflow with the circum-merger ISM that gives rise to the  radio and the X-rays.  
We used the public code PLUTO  \citep{Mignone2007}, with an HLL Riemann solver and a third order Runge Kutta time stepping.

The numerical simulations were done in two steps, spanning together over about ten orders of magnitude in length scale, both for the hydrodynamics and for the radiation. The hydrodynamics begins on a scale of $10^8$ cm where we inject the jet and it ends at  about $10^{18}$ cm, the location of the blast wave 1000 days (as measured in the observer frame) after the event. 

In each one of the relevant stages we  post-process the hydrodynamic results and calculate the emission (see Appendix for details). The prompt $\gamma$-ray emission upon the cocoon shock breakout,  takes place at a radius $\sim 10^{11}$ cm, where the emitting region width is about $10^{9}$ cm. The  macronova signal is produced at  $10^{15}-10^{16}$ cm, and the subsequent afterglow at  $10^{16}-10^{18}$ cm. 
As initial conditions we take the sub-relativistic ejecta that was ejected by the merger. The ejecta contained two components, a massive and slow  component ($0.05-0.1 {\rm~M_\odot}$ at 0.1-0.3c), which is inferred from the uv/optical/IR emission, and a fast low mass tail that is necessary, in our model,  for the breakout to take place at a large enough radius to produce the observed $\gamma$-rays.
This fast tail is not observed directly, but its  existence was predicted as an outcome of the shock that formed during the first collision between the two neutron stars  \citep{Kyutoku2014} and  it  was found  in early merger simulations (although with marginal resolution)  \citep{Hotokezaka2013}, where $\sim 10^{-6}-10^{-7}{\rm~M_\odot}$ has a velocity of 0.7-0.8c. \cite{Kiuchi2017} performed recently the highest resolution merger simulations ever and confirmed the existence of this fast component.   

\section{Results} 
\label{sec:results}
\begin{figure}
\includegraphics[width=0.5\textwidth]{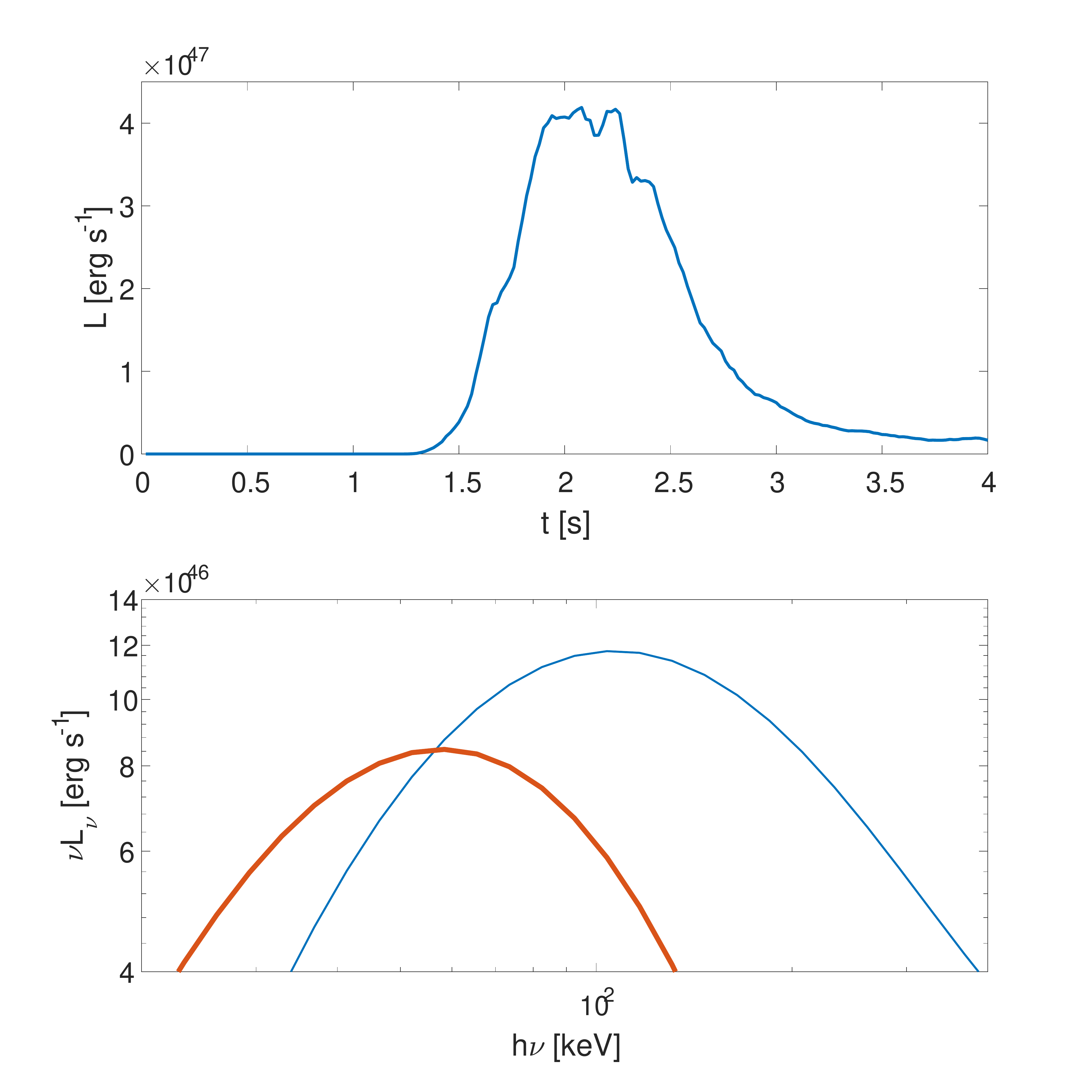}
\vskip -.5cm
\caption{The $\gamma$-ray light curve (top panel) and spectrum (bottom panel) during the shock breakout from the uncollimated choked jet simulation describe in the Appendix. The spectrum is divided to the emission during the peak (blue line) and during the tail (red line). The jet opening angle is 0.7 rad and it is fully choked long before the shock driven by the cocoon breaks out of the ejecta fast tail. The observer is at an angle of 0.6 rad. The $\gamma$-ray signal shows the main properties of GRB 170817A \citep{Goldstein17}. It starts rising about 1.5 s after the merger  having a peak that lasts about 0.5 s followed by a longer decay that lasts an additional second. The peak luminosity is comparable to that of GRB 170817A to within an order of magnitude (about a factor of 3 brighter). The spectrum shows a hard to soft evolution where the peak of $\nu F_\nu$ is $E_p \approx 110$ keV during the initial pulse (integrated up to 2.3 s) and $E_p \approx 55$ keV during the tail (after 2.3 s). This should be compared to GRB 170817A where in the initial pulse $E_p=185 \pm 62$ keV and  the tail's  spectrum is consistent with a blackbody with $T \approx 10$ keV, where $E_p \approx 40$ keV.}
\label{fig:gamma_rays}%
\end{figure}

\begin{figure}[h]
\includegraphics[width=0.5\textwidth]{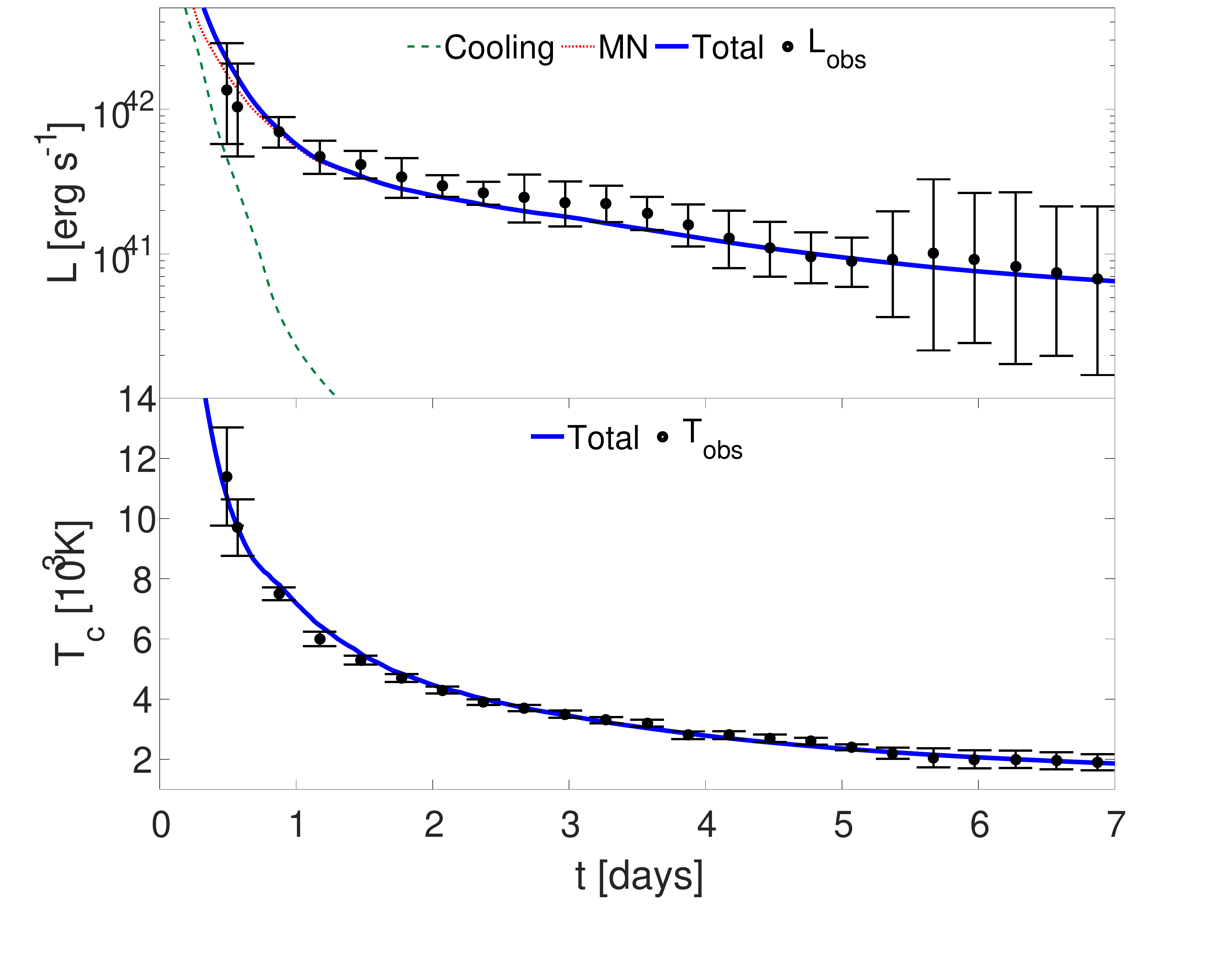}
\vskip -0.5cm
\caption{The optical light curve (top panel) and color temperature (bottom panel) from an uncollimated choked jet simulation  (the same simulation and same observing angle as in Fig. 1), see the Appendix for details. The data is taken from \citet{Kasliwal17}. Here we show the fit for a two component ejecta, where the opacity of slow moving material ($<0.1$c; mostly ejected along the equator) is $5 {\rm~cm^2/g}$, corresponds to a lanthanides rich material,  while that of fast moving material is $0.8 {\rm~cm^2/g}$, consistent with the expectation for a lanthanides poor material. A fit with a similar quality is obtained for a single lanthanides poor component outflow with a constant opacity of $0.8 {\rm~cm^2/g}$. The contribution of the two power sources are shown: The cocoon cooling emission in (dashed line) and the radioactive macronova (dotted line). The cocoon macronova emission is significant during the first day, after which the ejecta macronova dominates the emission.}
\label{fig:optical}%
\end{figure}

\begin{figure}
\includegraphics[width=0.5\textwidth]{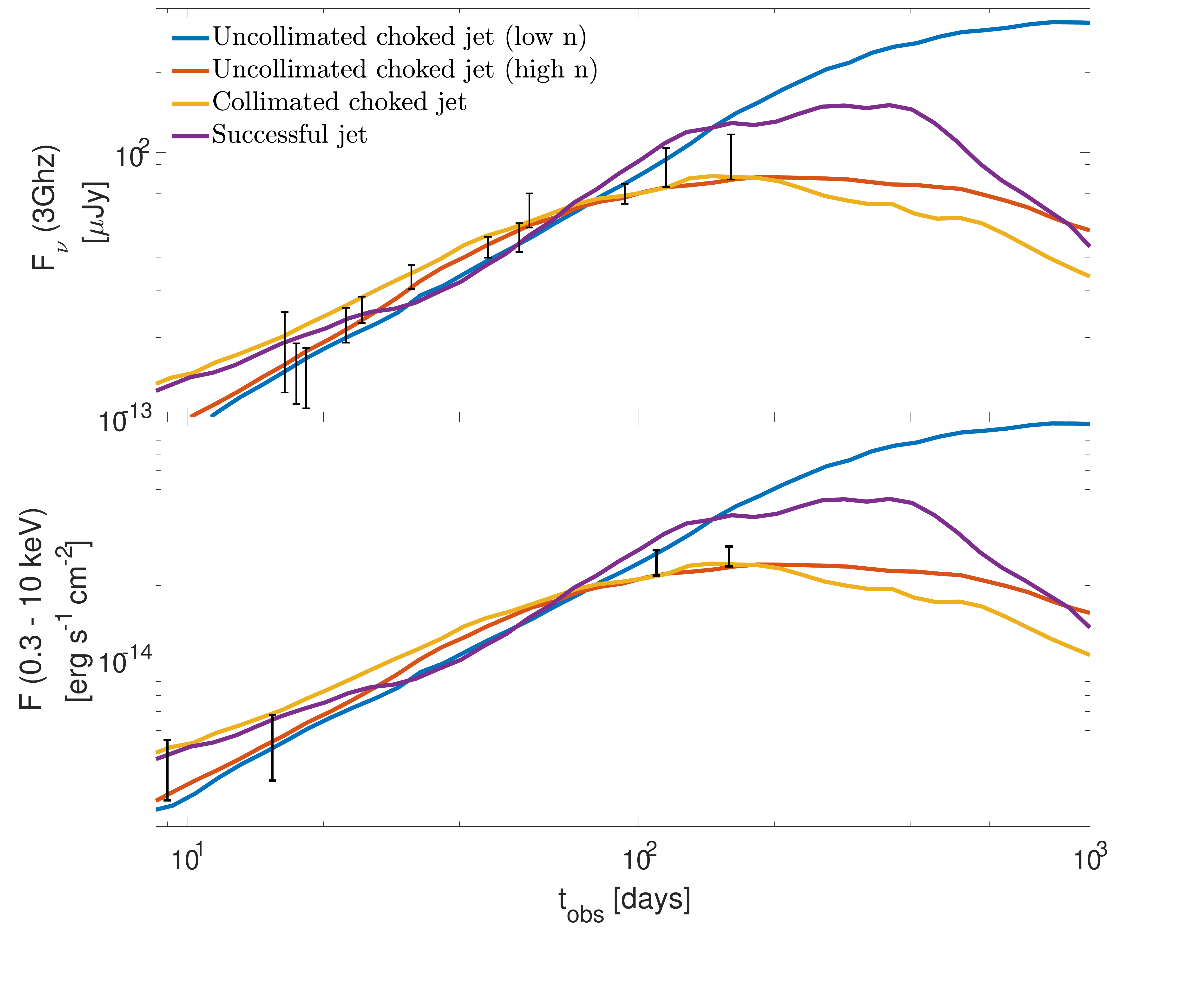}
\vskip -0.5cm
\caption{The afterglow emission compared with the radio and X-ray data \citep{Hallinan17,mooley2018,Troja17,margutti2018}. Shown are four simulations, two uncollimated, both with a choked jet and two collimated, one with a choked jet and one with a successful jet. In all simulations the observer is at 0.6 rad (see the Appendix). Both uncollimated choked jet simulations have the same outflow as  in figures \ref{fig:gamma_rays} and \ref{fig:optical}. The difference is that one has a high ISM density and a low magnetic field, 
($n= 0.05 {\rm~cm^{-3}}$, $\epsilon_B=10^{-5}$) and the other has a lower density and a higher magnetic field ($n= 5 \times 10^{-4} {\rm~cm^{-3}}$, $\epsilon_B=0.006$). The collimated choked jet afterglow simulation has $n= 0.02 {\rm~cm^{-3}}$, $\epsilon_B=7 \times 10^{-4}$ and the successful jet simulation has $n= 0.05 {\rm~cm^{-3}}$, $\epsilon_B=10^{-5}$. In all simulations $\epsilon_e=0.1$ and $p=2.17$. All models fit the observations very well and in all the cocoon dominates the emission during most (if not all) of the available observations. The models start to deviate only after about 100 days. In the collimated cases (both choked and successful) the emission near the peak contains a significant contribution from material along the jet axis (in the successful case it is the jet), and following the peak there is a fast decline. The afterglows of 	uncollimated choked jets also reach a peak, but the following decline is typically shallower than the collimated jets.  We note that the jet in the successful case is extremely energetic with an isotropic equivalent luminosity of $4.4 \times 10^{52}$ erg/s which is unusually high for sGRBs. This is the main reason that its contribution starts so late. } 
\label{fig:radio}%
\end{figure}

In the various simulations we varied the jet and ejecta properties. We focused on relatively wide jets with  opening angles $\sim 0.5-0.7$ rad. Such jets  are uncollimated and are fully choked in the ejecta dumping all their energy into the mildly relativistic cocoon. Only these jets can accelerate enough ejecta material to affect the early optical emission. Among those simulations we looked for those that fit the entire set of observations, finding several. An example of the emission from  such a simulation and  a comparison to the observations is shown in figures \ref{fig:gamma_rays}-\ref{fig:radio}. The jet in this simulation has a luminosity of $2.6 \times 10^{51}$ erg/s, an opening angle of 0.7 rad and it is launched for 1 s starting 0.8 s after the merger. The observer is set at an angle of 0.6 rad with respect to the jet symmetry axis. {In figure \ref{fig:radio} we show the afterglow for two different ISM densities, both fit current observations but each has a different future prediction.}

\begin{figure}
\includegraphics[width=0.5\textwidth]{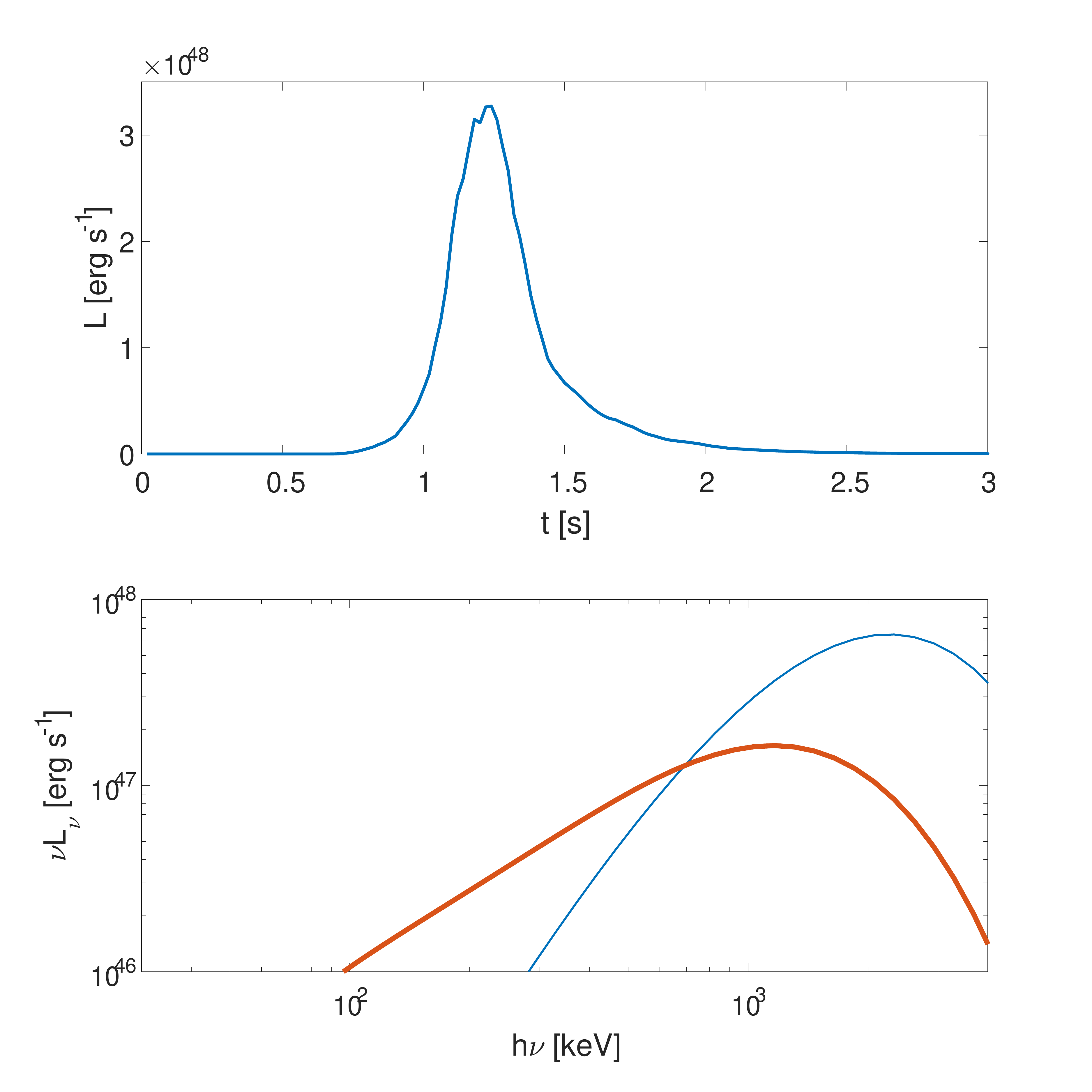}
\caption{The shock breakout $\gamma$-ray light curve (top panel) and spectrum (bottom panel) from the successful jet simulation. The blue thin line is the spectrum during the peak while the red thick line is the spectrum during the tail.}
\label{fig:gamma_rays_successful}%
\end{figure}

We also run two simulations of a narrow powerful jet with an initial opening angle of $10^o$  and a luminosity of $6.7 \times 10^{50}$ erg/s. In both simulations the jet and ejecta parameters are the same and the only difference is the duration over which the jet is launched. In one simulation the jet is launched for a duration that is long enough so it successfully breaks out of the ejecta and presumably produces a sGRB that can be observed along its axis.  {This successful jet with its cocoon, which is sometimes called "structured jet", is scenario E in figure 2 of \cite{mooley2018} and it is similar to the one considered by \cite{lazzati2017c} and \cite{margutti2018} }. In the second simulation the jet is launched for a short duration and it is choked before breaking out of the ejecta. In both cases the cocoon does not affect the uv/optical emission on a timescale of half a day, but we show that the shock breakout of the successful jet can potentially generate the $\gamma$-ray signal (figure \ref{fig:gamma_rays_successful}). In both simulations there are afterglow parameters (ISM density and microphysics) for which the observed radio and X-ray emission is produced  as shown in figure \ref{fig:radio}. 

\section{Was there a relativistic jet in GW170817?}
\label{sec:jet} 
A relativistic jet is necessary to produce a sGRB. Therefore, the question  whether a relativistic jet can successfully break out of a neutron star merger ejecta and in particular whether a jet broke out in this event  is interesting. In GW170817 we  consider the option of a successful jet as possible, but less likely. First, it doesn't explain the early optical light. Second, and more importantly, to explain the observed radio emission the jet should be either much weaker than the cocoon, so its contribution is unimportant, or it should be so powerful that its contribution begins only at very late time, at least months after the merger. Without fine tuning, the jet is not expected to have much less energy than the cocoon \citep{Nakar2017}, while a very powerful jet produces a very luminous sGRB and these are known to be extremely rare \citep{nakar2007,wanderman2015}. However, these are only circumstantial evidence.

Unfortunately a weak successful jet does not contribute to the observed emission and therefore we may  never know  if it existed or not. 
A powerful jet, however, 
might be detectable. If it is powerful enough it may be detectable while being still off-axis. In that case the radio light curve should rise   sharply, then decline as a power-law, after a short plateau. This unique prediction of the jet signature  is the easiest to identify. 
A  less powerful jet may contribute to the radio light curve, but its contribution may not be easily identified \citep[e.g., see figure \ref{fig:radio} and ][]{lazzati2017c,margutti2018}, since a continuous rise to a peak followed by a decay can arise either from  a successful or from a choked jet { \citep[see figure \ref{fig:radio} and][]{nakar2018a}. }

\begin{figure}	
\includegraphics[width=0.5\textwidth]{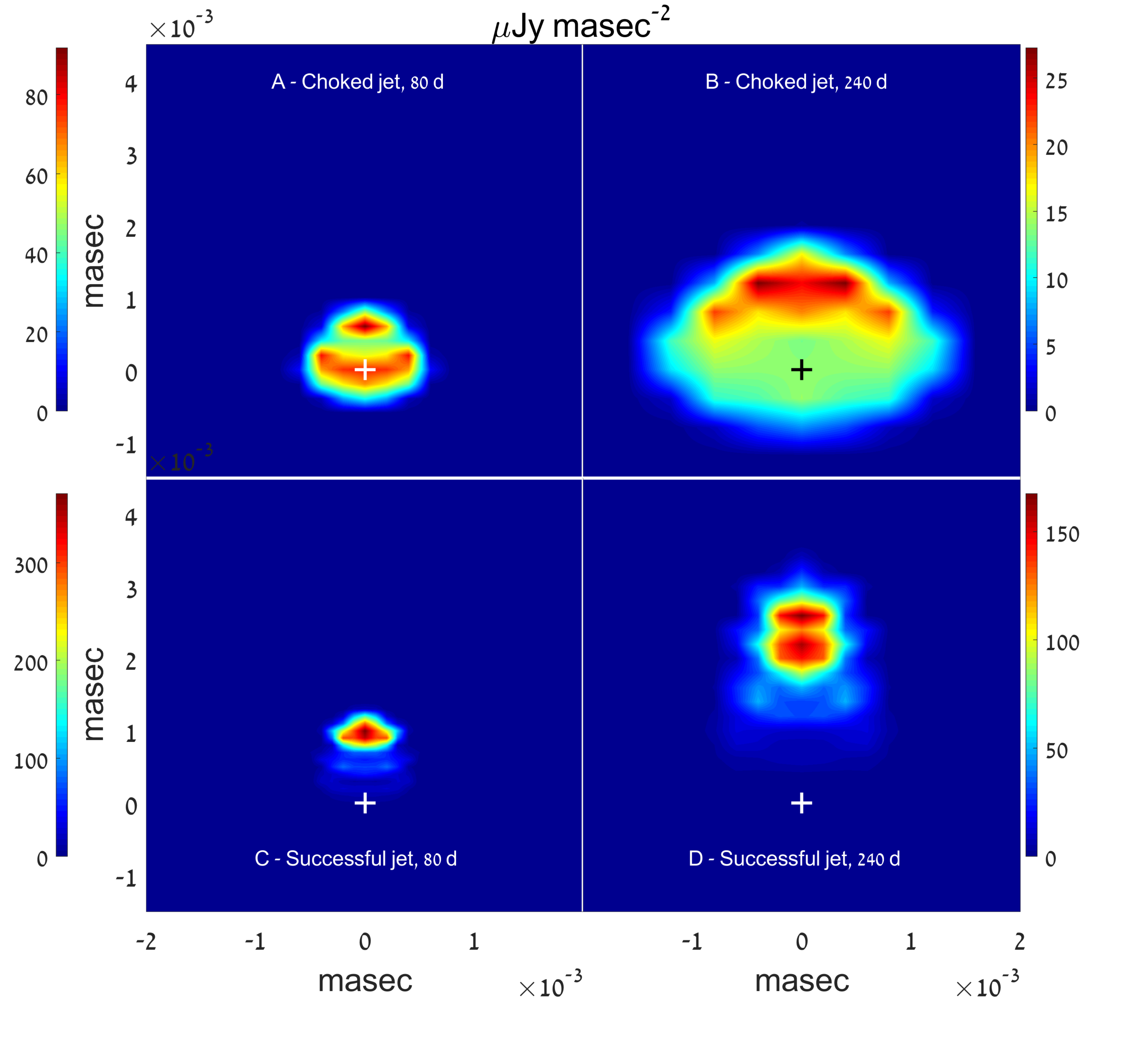}
\vskip -0.5cm
\caption{Radio images at 3GHz for two of the simulations presented in figure \ref{fig:radio}, the uncollimated choked jet with high density and the succesful jet. The images are taken on days 80 (panels A and C) and 240 (panels B and D), where the light curve of both afterglows peak. The crosses at [0,0] mark the line of sight. Both models show  a roughly linear growth of the image size with time, to a size that should be resolved by the VLBI. By day 80 the jet is already the main contributor in the successful jet model, whereas the wider cocoon leads to a bigger image in case of a choked jet. By day 240 both shapes remain similar but have grown bigger over time with the successful jet showing a more symmetric image with a centroid that is at a larger offset compared to the image on day 80. This is a general feature of successful jet images near the peak of the radio light curve, when the peak is dominated by the jet emission. We note that the image of the choked jet in this example is not general and in other simulations (that fit the radio light curve) we see a different image structures. The common property of most choked jet images is that they are asymmetric with one axis of the image being longer than the other.}  
\label{fig:images}%
\end{figure}

Additional information, especially in the radio near the time that the afterglow flux peaks, may provide the key to resolve this question. First, the radio image of the source is expected to be resolved soon. Figure \ref{fig:images} shows the images of a choked and a successful jet simulations. It shows that when there is a successful jet, the image becomes more symmetric and its centroid moves significantly (on a scale comparable to the image size), as the emission from the jet becomes more prominent. The jet dominates the emission once the light curve peaks and then the  motion of the image centroid stops. This evolution is generic and is expected to be seen in all the cases that a successful jet dominates the emission near the peak. The image of a choked jet is less predictive as it depends on the details of the cocoon structure and it varies between different simulations we carried out. However, its evolution is typically different than that of a successful jet. Generally, we found that it is much less symmetric near the peak of the light curve, and its centroid tend to move more slowly before the peak, than the image of a successful jet. 

Linear polarization may provide an additional valuable clue. The level of polarization depends on the geometry of the image and on the structure of the magnetic field. Therefore, the exact level of polarization at a given time will not give a definite answer, however the time evolution differ significantly between a choked and a successful jet.
A maximal polarization is seen when the source is moving at an angle of $1/\Gamma$ with respect to the line of sight \citep{sari1999}. Thus, if there is a dominant jet then the polarization is expected to rise significantly as the jet becomes more dominant, reaching a peak in the polarization near or slightly after the peak of the light curve. After the peak, as the flux starts its decay,  also the polarization level starts to  drop. In contrast, the polarization of a choked jet is not expected to show significant evolution before and near the peak of the light curve. To conclude, while there may be uncertainty in interpreting any one piece of the information carried by the afterglow light, when added altogether we have a powerful tool at hand to resolve the structure of the outflow.

\section{Conclusions} 
\label{sec:summary} 
In summary, we have shown that the entire set of electromagnetic observations that followed GW170817 can be explained by the different phases in the evolution of a mildly relativistic cocoon, formed during the propagation of a relativistic jet through the sub-relativistic ejecta. While the formation and evolution of the cocoon, as well as its importance for a signal observed away from the cone of the relativistic jet were predicted before GW170817 \citep{Nakar2017,lazzati2017a,Gottlieb17a},  
here we show that it is much more important than what was thought before.  
A clear prediction of this model is that future neutron star mergers will also show the signature of this mildly relativistic outflow. The current observations of GW170817 can be explained by a cocoon arising either from a choked jet or from a successful one, {where in the latter case the jet may significantly contribute to the emission at late times.}  While we favor the former, both are viable and GW170817 may or may not have produced a sGRB. Therefore, GW170817 is not yet a conclusive  evidence for the sGRB-NS merger connection. Nevertheless, it is certainly supporting this connection since it demonstrates that NS mergers most likely generate relativistic jets, even if in this specific event the jet was choked. This is especially true given that the duration distribution of sGRBs shows evidence for a large number of failed sGRBs where the jet is choked \citep{Moharana2017}.

This research was supported by the I-Core center of excellence of the CHE-ISF. OG and EN were partially supported by an ERC starting grant (GRB/SN) and an ISF grant (1277/13). TP was partially supported by an advanced ERC grant TReX and by a grant from the Templeton foundation. 


\appendix
\section{Numerical simulations}
\label{sec:nsimulation}

We carry out numerical relativistic hydro simulations of the jet ejecta interaction and of the resulting outflow and its interaction with the surrounding matter.  We then post-process the hydrodynamic results to calculate the observed emission. 
The hydrodynamics has four phases. First, the jet-ejecta interaction and the formation of the cocoon. 
Second, the breakout of the cocoon (and the jet if it is not choked) and its expansion and acceleration until it reaches a homologous expansion. Third, a phase of homologous expansion during which the interaction with the external medium is unimportant, and finally, once enough ISM material is collected, the slowing down of the ejecta due to the strong blast wave it drives into the surrounding ISM. 
The hydro simulations are carried out in two steps. In step I we calculate the jet propagation within the ejecta and the cocoon breakout and expansion. This step ends when the cocoon material reaches a homologous expansion. During the homologous expansion phase each element moves ballistically and cools adiabatically, and therefore we do not need to simulate this phase. We simply propagate the results from the end of phase I up to a radius that is smaller by about an order of magnitude than the radius at which the radio emission starts being significant. Then we start step II of the simulation that calculates the interaction of the outflow with the surrounding ISM. 

We used the public code PLUTO  \citep{Mignone2007}, with an HLL Riemann solver and a third order Runge Kutta time stepping. Throughout the simulations we apply an equation of state with a constant adiabatic index of $ 4/3 $, as appropriate for a radiation dominated and relativistic gas. We neglect gravity, as the gravitational dynamical times are longer than the typical interaction timescales at all phases of the simulations.

The hydro simulation begins at the time of the merger, that is $ t=0 $, with a radially expanding cold ejecta (core and tail). After a fraction of a second a relativistic jet is launched into the ejecta. The interaction of this jet with the ejecta forms a hot cocoon that    eventually breaks out of the ejecta (with a jet in the successful case) at $r_{bo}$. We let the relativistic outflow  reach $ 5-10 r_{bo} $ before terminating the simulation at $ 1.2\times 10^{12}\cm $.
We then stop the simulation and propagate the matter ballistically to a  radius of $ \sim 10^{16}\cm $ where we start it again to calculate the interaction with an external constant density medium.

Following the hydro simulations we post-process the data and calculate the observed radiation: the $\gamma$-rays  during the shock breakout phase;  the macronova's uv/optical/IR emission during the homologous expansion phase and   the radio and X-rays during the final stage of interaction with the ISM. 

We carried out several simulations of uncollimated choked jets where we varied various parameters, such as the exact ejecta mass and velocity profile, the maximal ejecta velocity and the jet luminosity and duration. We found several models where the resulting  signals resemble the electromagnetic counterparts of GW 170817. Here we describe one such simulation. 

Simulations of collimated jets are much more demanding computationally (see below). We  carried out, therefore, only two simulations, one of a successful jet and one of a choked jet. We found afterglow parameters for which the radio and X-rays from these simulations provide an good fit to the data (see figure \ref{fig:radio}). We calculated the $\gamma$-ray signal from the simulation of the successful jet and it is brighter and harder than GRB 170817A by about an order of magnitude (see figure \ref{fig:gamma_rays_successful}), but we expect that there are other parameters for which the shock breakout of a cocoon driven by a successful jet produces a $\gamma$-ray signal that is consistent with GRB 170817A.   

\subsection{Hydrodynamic simulations - Uncollimated choked jet}
\label{sec:choked_jet_setup}

We describe the details of the hydrodynamic simulation of the choked jet, for which we present the results in figures \ref{fig:gamma_rays}-\ref{fig:radio}. We use 2D simulations for all the phases. This is justified  since  the jet in this simulation is launched with a wide opening angle and therefore it is uncollimated. As a result (unlike the case of a collimated jet,  \citealt{Gottlieb17a}) the structure at the jet's head is not strongly affected by the symmetry axis   and the evolution in 2D and 3D is expected to be similar.

As initial conditions for part I we set the ejecta to be present from the base of the grid at $ r_{min} = 4\times 10^8 \cm$ up to $ r_{max} = 5.2 \times 10^9 \cm $. 
The ejecta is composed of two components, a massive and slow component, which we describe as the core, and a low-mass fast component, which we denote as the  tail. The core  extends from $ r_{min}$ to $ r_{max}/4 $, and the fast-tail from $ r_{max}/4 $ to $ r_{max} $. The core ejecta's density profile is:
\begin{equation}
\rho_{\rm{c}}(r,\theta) = \rho_0r^{-2}(\frac{1}{4}+\sin^3\theta)~,
\end{equation}
where $ \rho_0 $  is chosen for a total core ejecta mass $ M_c = 0.1\msun $.
Namely, the angular dependence is such that  3/4 of the ejecta mass near the equator at $1.0 \rad <\theta < \pi - 1.0 \rad $, where $\theta$ is the angle with respect to the jet axis.
The velocity profile of the  core is
\begin{equation}
v_{\rm{c}}(r) = v_{c,max}\frac{r}{r_c}~,
\end{equation}
where $ v_{c,max} = 0.2c $ is the core's maximal velocity. 
The normalisation of the fast-tail density  is chosen so its total mass is $ M_t = 5 \times 10^{-3}\msun $. 
With a steep power-law $\rho \propto v^{-14}$ and maximal velocity 0.7c. With this velocity profile only $ 10^{-8} M_\odot$ are moving at $v \ge  0.6c$. 
 
The jet is injected into the ejecta with a delay of $ 0.8\s $ for a total working time of $ 1\s $ and a total luminosity of $L_j=2.6 \times 10^{51} ~\rm{erg~ s^{-1}}$. The jet is injected with a specific enthalpy of 20 at and opening angle of $ 0.7\rad $ from a nozzle at the base of the grid with a size of $ 10^8\cm $.

We use a 2D cylindrical symmetric grid where the symmetry axis is $z$. For the grid of the first part of the simulation (from merger till the homologous phase) we use three patches along the $ x $-axis, the innermost one in the $x$-axis  resolves the jet's nozzle with 20 uniform cells from $ x = 0 $ to $ x = 2\times 10^8\cm $. The next patch stretches logarithmically from $ x = 2\times 10^8\cm $ to $ x = 2\times 10^{10}\cm $ with 800 cells, and the last patch has 1200 uniform cells to $ x = 1.2\times 10^{12}\cm $. In the $ z $-axis we employ two uniform patches, one from $ z_{min} = 4.5\times 10^8\cm $ to $ z = 2\times 10^{10}\cm $ with 800 cells, and the second to $ z = 1.2\times 10^{12}\cm $ with 1200 cells. In total the grid contains $ 2020 \times 2000 $ cells.

For the second part of simulation we propagate the last snapshot of the first simulation ballistically to a radius of $ \sim 10^{16}\cm $, where we let the relativistic outflow interact with the ISM. We verify that taking a smaller initial radius for this part does not affect the radio and X-ray light curve at the times of the observations.
The simulation can be scaled, so a single simulation can be used to calculate the hydrodynamic evolution in different ISM densities over a diffrent range of radii \citep{granot2012}. We present results for  
two such scalings, each with a different ISM number density and different radii $ R_1 $ and $ R_2 $ for the patches of the grid.
For the high ISM number density we use $ n = 5\times 10^{-2} \cm^{-3} $ with $ R_1 = 7\times 10^{15}\cm $ and $ R_2 = 4\times 10^{17}\cm $.
In the second simulation we use $ n = 5\times 10^{-4} \cm^{-3} $ with $ R_1 = 3.2\times 10^{16}\cm $ and $ R_2 = 1.8\times 10^{18}\cm $.
The grids are divided into two patches on each axis, the first of which are uniform with 500 cells on each axis, from $ 0 $ and to $ R_1 $ in the $ x $-axis and from $ z_{min} $ to $ R_1 $ in the $ z $-axis. This patch contains the entire grid of the final snapshot of part I  (propagated ballistically). The outer patches are identical with 2500 cells in each axis which are distributed logarithmically between $R_1$ and $ R_2 $. In total the grid has $ 3000 \times 3000 $ cells.

\subsection{Hydrodynamic simulation - Collimated successful and choked  jets}
\label{sec:successful_jet_setup}

We turn now to describe the details of the hydrodynamic simulation of the collimated jets. Here, the initial phase of the jet propagation within the ejecta core must be carried out using high resolution 3D simulation  \citep{Gottlieb17a}. This phase is very demanding computationally and therefore we carry out only two simulations. After the jet crosses the ejecta core and expands into the fast tail it is not collimated anymore and the evolution in 2D and 3D is expected be similar. We therefore take the result of the 3D simulation upon breakout of the jet from the core and map it to 2D by averaging over the azimuthal angle. We then continue the simulation in 2D.

We run two identical simulations except for a single difference, the jet launching time, so in one simulation the jet is choked while in the other it is successful. In both simulations at $ t = 0 $ a cold ejecta with $ M_c = 0.05\msun $ and $ v_{c,max} = 0.2c $ is present from $ r_{min} = 1.3\times 10^8\cm $ up to $ r_{max} = 3r_{min} $ with a radial density profile $\rho \propto r^{ -3.5}$. At $ t = 0.72\s $ a narrow jet with an opening angle of $ 10^\circ $ is injected into the system. The jet has a specific enthalpy of 20 and total luminosity $ 6.7 \times 10^{50}\rm{erg/s} $. In one simulation the jet is launched continuously for 1s and it breaks out of the ejecta successfully. In the second simulation the jet is launched for 0.4s and it is choked when its head is crossing most of the core ejecta. We stop the 3D simulations at t=1.4s, which is the time that the successful jet breaks out of the core ejecta and the choked jet is fully choked, and convert the 3D grid to 2D, while adding a homologous  fast-tail ejecta with a total mass $ M_t = 2\times 10^{-3}\msun $, a density profile $\rho \propto r^{-10}$ and velocity range of $ 0.2-0.8c$. The grids of the choked and successful jets scenarios is identical.

The grid of the 3D simulation is divided into two patches in each axis, where $z$ is the jet symmetry axis.
The inner patches are distributed uniformly in $x$ and $y$ (100 cells each) and $ z $ (400 cells) axes, extending to $ \pm 3\times 10^8\cm $ and $ 6\times 10^9\cm $, respectively. The $ z $-axis begins at $ z_{min} = 1.3\times 10^8\cm $. The second patches are logarithmic with 480 and 600 cells up to $\pm 9\times 10^{10}\cm $ in $x$ and $y$ and $ 1.2\times 10^{11} $ in $ z $. In total the 3D simulation contains $ 600 \times 600 \times 1000 $ cells.

After the 3D simulation ends its grid is mapped to a 2D grid with the exact same cell sizes in $x$ and $z$ and another patch is added to the grid in each axis stretching to $ 1.2\times 10^{12}\cm $ and $ 1.5\times 10^{12}\cm $ with 1200 and 1500 uniform cells in $ x $ and $ z $ axes, respectively. In total the simulation 2D contains $ 1490 \times 2560 $ cells and lasts $ 50\s $.

For part II we take the last snapshot of part I as the initial conditions and repeat the same procedure and cells distribution of the uncollimated simulation. Namely, we use different scalings for the successful and choked jet cases. For the former we use $ n = 5\times 10^{-2}\cm^{-3} $, $ R_1 = 1.3\times 10^{16}\cm $ and $ R_2 = 6.8\times 10^{18}\cm $. For the latter $ n = 2\times 10^{-2}\cm^{-3} $, $ R_1 = 10^{16}\cm $ and $ R_2 = 5.7\times 10^{18}\cm $.

\subsection{$\gamma$-ray emission}
\label{sec:gamma_calc}
The $\gamma$-ray emission following the shock breakout is calculated in the following way. First,  for each angle $ \theta $ we determine the time $ t_{bo}(\theta) $ and radius $ r_{bo}(\theta) $ at  which the shock breaks out from the ejecta at this angle. The luminosity is then calculated by finding  the photons that diffused to the photosphere during the time that passed since the shock crossing, approximating the diffusion as being radial. Specifically, at each lab frame time $t>t_{bo}$ and angle $ \theta $ we find the location $ r_0 $ from which the photons diffuse to the observer. Namely we find the location where the diffusion time to the photospheric radius, $ r_{ph}$ ($\tau(r_{ph}) = 1 $),  equals  the time that passed since  breakout: 
\begin{equation}
r_0 - r_{ph} = \frac{t-t_{bo}}{\Gamma(r_0)^2}\frac{c}{\tau(r_0)}~,
\end{equation}
where $ \Gamma(r_0) $ is the Lorentz factor of the emitting region. 
The energy, as measured in the fluid rest frame,  released between $t_1$ and  $t_2$ from a solid angle element $d \Omega$ is the total internal energy of the emitting region during this period:
\begin{equation}
dE'(t_1,t_2,\theta) = \int_{r_{0}(t_1)}^{r_0(t_2)} 4p(r)\Gamma(r) r^2 dr d\Omega \ .
\end{equation}
At the first time step we take $r_0(t_1)=\infty$. {To find the contribution to the observed energy, we  boost $ dE' $ from the rest frame of the photosphere to the observer frame, and take into account the light-travel time from $r_{ph}(\theta)$ to the observer. The total observed luminosity is found by integrating on all times at $t>t_{bo}$ over the solid angle. } 

The spectrum of the emitted radiation {at each time step from each angle } is approximated by a Wien spectrum, where the rest frame temperature is calculated using the procedure described in 
\cite{Gottlieb17b}. For each fluid element that releases the energy within a given time step we find temperature, T, for which the number density of photons that are generated within this fluid element during available time is $e/3kT$. We approximate this number by using the the pressure and the density of the element at the time that its energy is released and we solve equation 5 of  \cite{Gottlieb17b} {taking the available time as the maximum between the time that passed since the element was shocked and the time it takes the shock to cross the breakout layer (where $\tau= c/v_{sh}$).} If the temperature is higher than 50keV, then pairs are created, preventing the release of photons until the temperature drops to $\approx 50$ keV and  in such a case we set the temperature of the emitted radiation to 50 keV.

\subsection{UV/optical/IR emission}
\label{sec:UVOIR_calc}

We are interested in fitting the bolometric luminosity and temperature evolution of  the uv/optical/IR counterpart of GW 170817. We restrict the fit to the first week, after which  both the luminosity and the temperature are not well constraint. For the hydrodynamic evolution we use the final snapshot of part I of the choked jet hydrodynamical simulation, taken at time $t_f$. The homologous expansion after that time enables us to calculate the evolution of each fluid element and the adiabatic cooling of the radiation trapped in the fluid at any time $ t > t_f $. 

We estimate the opacity of the ejecta using two components, a low opacity component that corresponds to Lanthanides poor material and high opacity component that corresponds to Lanthanides rich material. This is motivated both by theoretical considerations that suggest that the ejecta at high latitudes is Lanthanides free while the outflow at low latitudes is Lanthanides rich \cite[e.g.][]{perego2014,kasen2015}, and more importantly, by the spectral evolution of the uv/optical/IR counterpart of GW 170817  \cite[e.g.][]{Evans17,smartt2017}. We find a good fit to the data  during the first week (figure \ref{fig:optical}) using $\kappa=0.8\ {\rm cm^2gr^{-1}}$ for  material moving at $v > 0.1c$, and $\kappa=5\ {\rm cm^2gr^{-1}}$ for  material moving at $v < 0.1c$ . This velocity criterion is applicable since the interaction with the jet accelerates the material at high latitudes to  $v > 0.1c$. We also test if a single component model can fit the data   \citep[e.g.][]{smartt2017} and find that a good fit (at least to the luminosity and temperature) can be obtained with a single component where all the ejecta have $\kappa \approx 0.8\ {\rm cm^2gr^{-1}}$.

The emission is calculated assuming that the photons diffuse radially. For each angle $ \theta $, we  calculate first  the optical depth to infinity, $\tau(r,\theta)$. We determine the trapping radius $ r_t(\theta) $ where $ \tau(r_t(\theta)) = c/v $, above which photons diffuse freely to the observer, and the photosphere $ r_{ph}(\theta) $ in which $ \tau(\theta) = 1 $. 

The emission is powered by two sources: the diffusion of the internal energy (cooling emission) and the radioactive heating (macronova). For the cooling emission we find the rest frame energy flux at $ r_t(t,\theta) $, which has been trapped up until the current time $ t $.
\begin{equation}
\Phi(t,\theta)_{cool} = 4p_tr_t(\theta)^2v_td\Omega~,
\end{equation} 
where $ p_t $ and $ v_t $ are the pressure and velocity at $ r_t(\theta) $.

We approximate the macronova luminosity as coming from the instantaneous heating rate at $ r > r_t $:
\begin{equation}
\Phi(t,\theta)_{MN} = \int_{r_t(\theta)}^\infty \dot{\epsilon} \rho(r)\Gamma(r) r^2 dr d\Omega~,
\end{equation}
where $ \rho $ and $ \Gamma $ are the density and Lorentz factor of the emitting region and 
the radioactive heating rate per unit of mass is approximated as \citep[e.g.][]{hotokezaka2016}
\begin{equation}
\dot{\epsilon} \approx \dot{\epsilon}_0(f_\gamma(t) +0.5f_e(t))(\frac{t}{\rm{day}})^{-\alpha},
\end{equation}
where $\dot{\epsilon}_0 \sim 10^{10}~\rm{erg~gr^{-1}~s^{-1}}$ and $\alpha \approx -1.1$ --- $-1.4$. $f_\gamma$ and $f_e$ are the fraction of the energy in gamma-rays and electrons that is deposited in the ejecta, respectively. As we are interested 
in the emission during the first week we approximate $f_e=1$ and $f_\gamma =1-e^{-(\frac{t_0}{t})^2} $, where $t_0 \sim 1$ d is the typical gamma-rays escape time. For the fit presented in figure \ref{fig:optical} we use 
$\dot{\epsilon}_0 = 2 \times 10^{10}~\rm{erg~gr^{-1}~s^{-1}}$, 
$\alpha=1.1$ and $t_0=0.7$ d. 

Along each angle we approximate the emission at any time as being emitted isotropically at the photospheric radius, $ r_{ph}(\theta) $, with a luminosity $\Phi(t,\theta)_{cool}+\Phi(t,\theta)_{MN}$ in the photosphere rest frame. We alos approximate the emission as being in thermal equilibrium at the photosphere, so emitted spectrum is a blackbody with a rest frame temperature that corresponds to the luminosity and $ r_{ph}(\theta) $. Finally, similarly to \ref{sec:gamma_calc} we integrate and boost the luminosity and temperature to the observer's reference frame. The spectrum that we obtain from the integration is not a blackbody and the temperature that we present is the color temperaure.

\subsection{Radio and X-ray afterglow emission}
\label{sec:radio_calc}

We model the afterglow emission by the synchrotron radiation of relativistic electrons accelerated in the shocks formed between that cocoon and the ISM. We use the standard afterglow modelling that parameterizes the microphysics using a constant fraction of the internal energy that goes to electrons, $\epsilon_e$,  a constant fraction of the internal energy that goes to the magnetic field, $\epsilon_B$, and a power-law distribution for the accelerated electrons with index $p$.  Given the low density of the ISM  the radio and the X-ray bands are not affected by cooling or by absorption. In fact both bands are on a single power-law segment above the typical synchrotron and self absorption frequencies and below the cooling frequency. Therefore the emission from each fluid element is not affected by its history (i.e., cooling) or by the conditions in other fluid elements (i.e., absorption). 

Using the microphysical  parameterization and the pressure and the density in each cell at each time-step we find the electron distribution and the magnetic field  and calculate the rest frame synchrotron emissivity using standard afterglow theory \citep[e.g.][]{sari98}. Then to find the observed luminosity at each observing angle we integrate over equal arrival time surfaces taking the proper Lorentz boost from each element.

Since the radio and the X-rays are on the same spectral power-law index, $-(p-1)/2$, the observations set  $p=2.17$. We also set $\epsilon_e=0.1$. The free parameters we have are external density, $n$,  and $\epsilon_B$. The values we use to fit the data in the various simulations (which are not unique) are given in figure \ref{fig:radio}.

\end{document}